\documentstyle[12pt]{article}
\begin{document}\begin{flushright}\thispagestyle{empty}
OUT--4102--73\\
hep-th/9805025\\
8 May 1998                      \end{flushright}\vspace*{2mm}\begin{center}{
                                                       \Large\bf
A dilogarithmic 3-dimensional Ising tetrahedron        }\vglue 10mm
                                                       {\large{\bf
D.~J.~Broadhurst                                       $^{1)}$}\vglue 4mm
Physics Department, Open University                    \\[3pt]
Milton Keynes MK7 6AA, UK     }\end{center}\vfill\noindent{\bf Abstract}\quad
In 3 dimensions, the Ising model is in the same universality class as
$\phi^4$-theory, whose massive 3-loop tetrahedral diagram, $C^{Tet}$, was of
an unknown analytical nature. In contrast, all single-scale 4-dimensional
tetrahedra were reduced, in hep-th/9803091, to special values of
exponentially convergent polylogarithms. Combining dispersion relations with
the integer-relation finder PSLQ, we find that
$C^{Tet}/2^{5/2}={\rm Cl}_2(4\alpha)-{\rm Cl}_2(2\alpha)$, with
${\rm Cl}_2(\theta):=\sum_{n>0}\sin(n\theta)/n^2$ and $\alpha:=\arcsin\frac13$.
This empirical relation has been checked at 1,000-digit precision and readily
yields 50,000 digits of $C^{Tet}$, after transformation to an exponentially
convergent sum, akin to those studied in math.CA/9803067. It appears
that this 3-dimensional result entails a polylogarithmic ladder beginning with
the classical formula for $\pi/\sqrt2$, in the manner that 4-dimensional
results build on that for $\pi/\sqrt3$.
\vfill\footnoterule\noindent
$^1$) D.Broadhurst@open.ac.uk;
http://physics.open.ac.uk/$\;\widetilde{}$dbroadhu
\newpage\setcounter{page}{1}
\newcommand{\df}[2]{\mbox{$\frac{#1}{#2}$}}
\def\bk{\mbox{\bf k}}

\section{Introduction}

In 3 dimensions, the universality class of the Ising model
includes $\phi^4$ theory, which entails at the 3-loop level
a tetrahedral Feynman diagram, corresponding to the
symmetrical 9-dimensional integral~\cite{GKM}
\begin{equation}
C^{Tet}:=\frac{1}{\pi^6}\int d^3\bk_1d^3\bk_2d^3\bk_3\Delta(\bk_1)\Delta(
\bk_2)\Delta(\bk_3)\Delta(\bk_1-\bk_2)\Delta(\bk_2-\bk_3)\Delta(\bk_3-\bk_1)
\end{equation}
with $\Delta(\bk):=1/(|\bk|^2+1)$ as the unit-mass propagator.
A numerical value, $C^{Tet}\approx0.1739006$, was obtained in~\cite{NMB}
and checked in~\cite{GKM,AKR}.
We shall show that the dispersive methods of~\cite{mas,sixth} enable a
reduction of $C^{Tet}$,
as for any assignment of masses, to single integrals of
logarithms. Then we shall describe how the lattice algorithm PSLQ~\cite{PSLQ}
achieved a very simple reduction of $C^{Tet}$ to a
Clausen integral, which gives an exponentially convergent sum
that reveals a new feature of the distinctive mapping~\cite{DK} of
diagrams~\cite{sixth,BKP,BDK,BGK,BK15,BK4} to
numbers~\cite{Eul,BBB,poly,BBBL} provided by quantum field
theory.

\section{Dispersive integral}

Let $C(a,b)$ be the tetrahedron with non-adjacent lines carrying masses
$a$ and $b$, while the other 4 lines retain unit mass. Then a long
dispersive calculation produces a short
result:
\begin{equation}
C(a,b)=-\frac{16}{b}\int_{2}^{\infty}{dw\over(w+a)D(w,b)}~{\rm arctanh}
\left({N(w,b)\over D(w,b)}\right)\label{cab}
\end{equation}
where the denominator function
\begin{equation}
D(w,b):=w\sqrt{w^2+b^2-4}\label{Dwb}
\end{equation}
is regular at the 2-particle threshold, $w=2$, provided that $b>0$, and
\begin{eqnarray}
N(w,b)&=&w^2-2(2+b)\mbox{ ~for~ }w\in[2,2+b]\label{N2}\\
N(w,b)&=&\qquad~w~b\qquad\mbox{ ~for~ }w\in[2+b,\infty]\label{N3}
\end{eqnarray}
specify a numerator that is continuous in value, though not in derivative,
at the 3-particle threshold, $w=2+b$.
The origins of~(\ref{cab}--\ref{N3}) will be outlined,
neglecting factors of 2 and $\pi$.
\begin{enumerate}
\item
Let $I(\bk,b)$ be the 2-point function obtained by cutting the
tetrahedron at the line with mass $a$, so that
\begin{equation}
C(a,b)\sim\int\frac{d^3\bk}{|\bk|^2+a^2}\,I(\bk,b)\label{int1}
\end{equation}
with the 2-point function given by a dispersion relation of the form
\begin{equation}
I(\bk,b)\sim\int_2^\infty\frac{w\,dw}{w^2+|\bk|^2}\,\sigma(w,b)\label{int2}
\end{equation}
where $\sigma$ is the spectral density of $I$, considered
in 2+1 spacetime dimensions.
We perform this anti-Wick rotation, away from the 3 spatial dimensions of
condensed matter, in order to exploit the Cutkosky rules of Minkowski-space
quantum field theory, as in~\cite{mas}.
An interchange of order of integration
in~(\ref{int1},\ref{int2}) gives
\begin{equation}
C(a,b)\sim\int_2^\infty\frac{w\,dw}{w+a}\,\sigma(w,b)
\end{equation}
which explains the simple dependence on $a$ of the integrand
in~(\ref{cab}).
\item
The spectral density
\begin{equation}
\sigma(w,b)=\theta(w-2)\sigma_2(w,b)+\theta(w-2-b)\sigma_3(w,b)\label{s23}
\end{equation}
receives contributions from intermediate states with 2 and 3 particles.
In the first case, $\sigma_2(w,b)\sim\Re F(w+i0,b)/w$
entails a 1-loop form factor, $F$.
This may also be calculated dispersively, from its imaginary part
\begin{equation}
\Im F(w+i0,b)\sim
\frac{1}{w}\int_0^\pi\frac{d\phi}{2k^2(1-\cos\phi)+b^2}
=\frac{\pi}{w\,b\sqrt{w^2+b^2-4}}
\end{equation}
where $k:=\sqrt{(w/2)^2-1}$ and $\phi$
are the centre-of-mass 2-momentum and scattering angle,
in the elastic scattering of unit-mass particles,
by exchange of a
particle of mass $b$, in 2+1 spacetime dimensions.
This is the origin of the square root in~(\ref{Dwb}).
\item
It is now straightforward to calculate
\begin{equation}
w\,b\,\sigma_2(w,b)\sim\Re\int_2^\infty\frac
{x\,dx}{x^2-w^2+i0}\,\frac{1}{D(x,b)}
\end{equation}
and obtain logarithms from the real part
of the form factor. Maple produced 3 arctanh functions,
which were combined, by hand, to give the numerator~(\ref{N2}).
\item
The 3-particle intermediate state yields the Dalitz-plot integral
\begin{equation}
\sigma_3(w,b)\sim\Re
\int_{b(2+b)}^{w(w-2)}\frac{ds}{s}
\int\frac{dt}{t}\,\frac{1}{\sqrt{J(s,t,w^2,b^2)}}
=\int_{b(2+b)}^{w(w-2)}\frac{ds}{s}\,\frac{\pi}{\sqrt{-J(s,0,w^2,b^2)}}
\end{equation}
where $s$ and $t$ are the denominators of the propagators of the
two particles that are still
off-shell and the $t$ integration is over the range in which
the Jacobian
\begin{equation}
J(s,t,u,v):=-(s~t-u~v)(s+t+4-u-v)-(s-t)^2\,\label{J}
\end{equation}
is positive. Maple produced 2 arctanh functions,
to be added to the 3 from $\sigma_2$. Manual combination of these 5
logs produced the amazingly simple numerator~(\ref{N3}).
\end{enumerate}
This method is clearly generalizable to give a single integral
of logs in any mass case.

\section{Superconvergence and KLN cancellations}

The factor $-16/b$ in~(\ref{cab}) looks alarming, at first sight.
The integral is manifestly finite as $a\to0$. Field theory
proves that $C(a,b)=C(b,a)$, notwithstanding the very
different ways that the masses $a$ and $b$ enter the integral.
Hence $C(a,b)$ is finite as $b\to0$, despite the factor of
$1/b$. Already we see that potentially linear infra-red divergences
have been cancelled, by combining 2-particle and 3-particle
intermediate states in~(\ref{N3}). This parallels the 4-dimensional
cancellation of logarithmic divergences, from virtual and real
soft photons, by the Kinoshita--Lee--Nauenberg mechanism~\cite{KLN}.
However, it is still not safe to take the limit $b\to0$, blithely,
since the contributions from $w>2+b$ are manifestly negative,
and have a $1/(w-2)$ singularity as $b\to0$.

The key to handling this tricky limit is the superconvergence
relation
\begin{equation}
0=\int_2^\infty{dw\over D(w,b)}~{\rm arctanh}
\left({N(w,b)\over D(w,b)}\right)\label{super}
\end{equation}
which ensures that $\lim_{a\to\infty}a\,C(a,b)=0$. Thus one may make
the replacement
\begin{equation}
{1\over w+a}\,\longrightarrow\,{1\over w+a}-{1\over 2+a}
\,=\,-{w-2\over(w+a)(2+a)}
\end{equation}
in~(\ref{cab}).
Then the factor $w-2$ suppresses the singularity at
threshold in the limit $b\to0$, giving the elementary integral
\begin{equation}
C(a,0)=\frac{16}{2+a}\int_2^\infty\frac{dw}{w(w+a)(w+2)}
={16\log(1+a/2)-8a\log2\over4a-a^3}
\end{equation}
in agreement with a more general case, given in~\cite{AKR}.
The values
\begin{eqnarray}
C(0,0)&=&2-\log4\\
C(1,0)&=&\df83\log\df98\\
C(2,0)&=&\log2-\df12\label{c20}\\
C(4,0)&=&\df13\log\df43\\
C(6,6)&=&\df{1}{12}\log2
\end{eqnarray}
entail only $\log2$ and $\log3$. This observation prompted the next step.

\section{Dilogarithms at $b=2$}

By giving numerical evaluations to the
lattice algorithm PSLQ, it was discovered that
$C(a,2)$ evaluates to dilogs with simple rational arguments,
for $a\in\{1,2,4,6\}$, namely
\begin{eqnarray}
C(1,2)&=&\pi^2+4{\rm Li}_2(\df{1}{16})-8{\rm Li}_2(\df16)-16{\rm Li}_2(\df14)
-2\log^23-4\log^22\\
C(2,2)&=&\df{\pi^2}{12}-{\rm Li}_2(\df14)-\log^22\\
C(4,2)&=&\df38{\rm Li}_2(\df14)+\df18\log^23-\df34\log2\log\df32\\
C(6,2)&=&\df29{\rm Li}_2(\df14)-\df19{\rm Li}(\df{1}{16})-\df{1}{18}\log^22
\end{eqnarray}
which indicated a dilogarithmic dependence of $C(a,2)$ on $a$.
Combining the superconvergence relation
with the simplicity of $D(w,2)=w^2$, a lengthy
expression was proven by computer algebra, and then simplified by hand
to give
\begin{eqnarray}\df14a^2C(a,2)&=&
3{\rm Li}_2(a/(a+2))-2{\rm Li}_2(a/(2a+4))
+{\rm Li}_2(2a/(a-2))-{\rm Li}_2(a/(a-2))\nonumber\\&&{}
+2{\rm Li}_2(-a/4)+\log^2(1+a/2)-\log(1-a^2/4)\log2
\end{eqnarray}
which shows that $C(0,2)=\log2-\df12$, in agreement with~(\ref{c20}).
Thanks to advice from Arttu Rajantie, it became clear that the 5
dilogs could be simplified to give 2, using transformations of
${\rm Li}_2(x):=-\int_0^x(dy/y)\log(1-y)$.
The most compact formula is
\begin{equation}
\df14a^2C(a,2)={\rm Li}_2((a-2)/(a+2))-2{\rm Li}_2(-2/(a+2))
-\df{1}{12}\pi^2\,.
\end{equation}

\section{PSLQ and the symmetric tetrahedron}

The previous results suggested the hypothesis that
the totally symmetric tetrahedron, $C^{Tet}:=C(1,1)$,
is a dilogarithm.
With the help of PSLQ, it was eventually
reduced to a Clausen integral of startling simplicity:
\begin{equation}
\frac{C(1,1)}{2^{5/2}}=-\int_{2\alpha}^{4\alpha}
d\theta\log(2\sin\df12\theta)\label{cl2}
\end{equation}
with $\alpha:=\arcsin\df13$.
A proof appears to be rather difficult,
though~(\ref{cl2}) has been confirmed numerically, at 1,000-digit precision.
The discovery route was typical of work with PSLQ. Splitting $C(1,1)$
into contributions below and above the 3-particle threshold,
one finds that the latter involve terms of the form
$\sqrt2\,{\rm Cl}_2(j\alpha+k\pi/6)$, with
\begin{equation}
{\rm Cl}_2(\theta):=\Im{\rm Li}_2(\exp(i\theta))
=\sum_{n>0}\frac{\sin(n\theta)}{n^2}
\end{equation}
and integer values of $j$ and $k$.
There appeared to be little prospect of reducing all terms
to this set of constants, by analytical methods alone. Yet PSLQ found
that the total is so reducible and also found many relations between
such Clausen values and the constants
$\{\pi\log2,\pi\log3,\alpha\log2,\alpha\log3\}$.
As so often remarked in field theory, the whole:
\begin{equation}
{C(1,1)\over2^{5/2}}={\rm Cl}_2(4\alpha)-{\rm Cl}_2(2\alpha)\label{ans}
\end{equation}
turned out to be far simpler than its parts.
As a final bonus, this was transformed, again with the aid of PSLQ,
to the exponentially convergent sum
\begin{equation}
C(1,1)=\sum_{n=0}^\infty\frac{(-1/2)^{3n}}{n+\frac12}
\left(\frac{1}{n+\frac12}-3\log2-\sum_{m=1}^n\frac{3}{m}\right)\label{exp}
\end{equation}
formed from terms found in integer relations with
$\sqrt2{\rm Cl}_2(j\alpha+k\pi/6)$.
This last result enables rapid computation in a single do-loop.
The first 50 digits of
\begin{equation}C^{Tet}:=C(1,1)={\tt
0.17390061066200274272650601711566596761380833829869
}\end{equation}
result in a trice, with 50,000 digits taking only 40 minutes on a 233 MHz
Pentium. The first 1,000 digits agree with numerical quadrature
of dispersive integrals, generously undertaken by Greg Fee, at CECM.

After this work was completed, Arttu Rajantie drew attention to an
alternative representation of massive 3-dimensional tetrahedra~\cite{AKR},
obtained by the method of differential equations~\cite{AVK}.
In the totally symmetric case this gives~\cite{AKR}
\begin{equation}
{C(1,1)\over2^{5/2}}=
\int_0^1\frac{dx}{\sqrt{3-x^2}}\,\left(\log\frac34+\log\frac{3+x}{2+x}
-\frac{x^2}{4-x^2}\log\frac{4}{2+x}+\frac{x}{2+x}\log\frac{3+x}{3}\right)
\end{equation}
which appears to be no easier to reduce to~(\ref{ans}) than
the dispersive integral~(\ref{cab}).

\section{Conclusions}

Thus PSLQ has shown that I was off target
when suggesting at the recent Rheinsberg workshop
that a super-renormalizable theory~\cite{GKM,AKR}
might be less interesting, mathematically, than QCD~\cite{sixth}.
In fact, the Ising tetrahedron is as intriguing as those in QCD.

One now sees that the symmetric 3-dimensional tetrahedron
is given by~(\ref{exp}) as an exponentially convergent sum
that sits close to the classical formula~\cite{BBP}
\begin{equation}
{\pi\over\sqrt2}=\sum_{n\ge0}\frac{(-1/2)^n+(-1/2)^{3n+2}}{n+\df12}\,.
\label{b2}
\end{equation}
This association resonates strongly with the recent reduction~\cite{sixth}
of a 4-dimensional tetrahedron, in the 3-loop QCD corrections to the
electro-weak
rho-parameter~\cite{rho,rhop},
to a sum of squares of two distinguished dilogarithms, namely $\zeta(2)$
and ${\rm Cl}_2(\pi/3)$. The latter was first encountered in 1-loop
massless 3-point functions~\cite{CG} and then
in the pioneering work of van der Bij and Veltman~\cite{BV}
on 2-loop massive diagrams.
In the massive case it appears in association with
\begin{equation}
{\pi\over\sqrt3}=\sum_{n\ge0}\frac{(-1/3)^n}{n+\df12}\,.\label{b3}
\end{equation}
It remains to be seen whether the `magic' connection proven
in~\cite{DT}, between massless and massive instances of
${\rm Cl}_2(\pi/3)$, is generalizable to the quadrilogarithms
found in~\cite{sixth} or to the dilogarithm~(\ref{ans}) found here.

In conclusion: 3-loop single-scale vacuum diagrams in
4 dimensions~\cite{sixth}
evaluate to quadrilogarithms of the sixth root of unit,
$\exp(i\pi/3)=(1+i\sqrt3)/2$, while in 3 dimensions we have now encountered
dilogarithms of $\exp(i\alpha)=(\sqrt8+i)/3$. In both cases,
there are remarkable transformations to exponentially convergent sums.
In the 4-dimensional case, these entail polylogarithmic ladders,
akin to those in~\cite{poly}, beginning with~(\ref{b3});
in 3 dimensions~(\ref{b2}) appears to provide
the lowest rung. In both cases, the results are of a simplicity, scarcely
to be expected from the method, that was revealed by PSLQ~\cite{PSLQ}.

\newpage\noindent{\bf Acknowledgements} I thank David Bailey and Greg Fee,
for computational assistance, Jochum van der Bij,
Andrei Davydychev, Dirk Kreimer, Gernot M\"unster, Willi van Neerven,
Arttu Rajantie and Bas Tausk, for advice,
and Johannes Bl\"umlein, Fred Jegerlehner and Tord Riemann, for
hospitality at Zeuthen. As so often, Dirk Kreimer
provided the vital stimulus.

\raggedright

\end{document}